\documentclass[sigconf]{acmart}

\usepackage{booktabs} 
\usepackage{bm}
\usepackage{algorithmic}
\usepackage{graphicx}
\usepackage{textcomp}
\usepackage{xcolor}
\usepackage{multirow}
\usepackage[noend, linesnumbered, ruled, lined]{algorithm2e}
\usepackage{verbatim}
\usepackage{adjustbox}
\usepackage{diagbox}
\usepackage[normalem]{ulem}
\usepackage{pifont}
\usepackage{tabularx}
\usepackage[most]{tcolorbox}

\usepackage{booktabs}
\usepackage{bm}
\usepackage[table]{xcolor}
\usepackage{booktabs}
\usepackage{tabularx}
\usepackage{array}
\usepackage{booktabs}
\usepackage{tabularx}
\usepackage{array}
\usepackage{algorithmic}
\usepackage{graphicx}
\usepackage{textcomp}
\usepackage{xcolor}
\usepackage{subfigure}
\usepackage{multirow}
\usepackage[noend, linesnumbered, ruled, lined]{algorithm2e}
\usepackage{hyperref}
\usepackage{verbatim}
\usepackage{adjustbox}
\usepackage{diagbox}
\usepackage{multirow}
\usepackage[normalem]{ulem}
\usepackage{ulem}
\usepackage{pifont}
\usepackage{tabularx}
\setcopyright{rightsretained}
\usepackage{dblfloatfix}
\usepackage{enumitem}
\usepackage{fontawesome5}
\renewcommand\footnotetextcopyrightpermission[1]{}
\setcopyright{none}
\usepackage{booktabs}
\usepackage{tabularx}
\usepackage{array}

\begin{document}
\title{OneBar: An End-to-End Content-Grounded Generative Query Recommendation Framework for E-Commerce Video Feeds}


\author{%
\texorpdfstring{%
\begin{tabular}{@{}c@{}}
\textbf{Yao Tang\textsuperscript{1}\textsuperscript{*}},
\textbf{Ying Yang\textsuperscript{2}\textsuperscript{*}},
\textbf{Ben Chen\textsuperscript{2}\textsuperscript{*}\textsuperscript{\dag}},
\textbf{Yufei Ma\textsuperscript{2}},
\textbf{Zihan Liang\textsuperscript{2}},
\textbf{Chenyi Lei\textsuperscript{2}}\\[2pt]
\textbf{Wenwu Ou\textsuperscript{2}},
\textbf{Jian Liu\textsuperscript{1}}\\[5pt]
\textsuperscript{1}Zhejiang University,
\textsuperscript{2}Kuaishou Technology\\[5pt]
{\small\faIcon{envelope}}\quad \texttt{benchen4395@gmail.com}
\end{tabular}%
}{%
Yao Tang, Ying Yang, Ben Chen, Yufei Ma, Zihan Liang, Chenyi Lei, Wenwu Ou, and Jian Liu
}%
}

\thanks{$^{*}$Equal contribution.} 
\thanks{$^{\dagger}$Corresponding author.}

\renewcommand{\shortauthors}{Chen et al.}







\begin{abstract}

Short-video platforms now expose clickable search entries beneath the video player, enabling users to easily express content-induced search intent. However, conventional query recommendation systems on short-video platforms suffer from latency constraints and objective misalignment, while recent generative approaches struggle with noisy content-side metadata and preference drift. To address these issues, we propose OneBar, an end-to-end generative framework for real-time query recommendation for
E-Commerce video feeds. OneBar features three key innovations: (1) a collaborative-multimodal intent grounding module that fuses multimodal video understanding and behavior-derived collaborative anchors; (2) a Unified End-to-End architecture equipped with a prompt-compression mechanism for efficient online serving; and (3) a progressive preference learning strategy for efficient preference-internalization, which internalizes hierarchical behavior preferences into the generative policy, eliminating the need for a separately trained reward model. Compared with online base, OneBar increases Query Exposure by 16.91\% and Query Click by 18.68\%, while maintaining a slight Query CTR gain of 0.19\%. The additional search traffic further contributes to 20.36\% more guided orders and 21.67\% higher GMV. 

\end{abstract}



\maketitle

\section{Introduction}
\label{sec:introduction}

\begin{figure}[!t]
    \centering
    \includegraphics[width=0.95\linewidth]{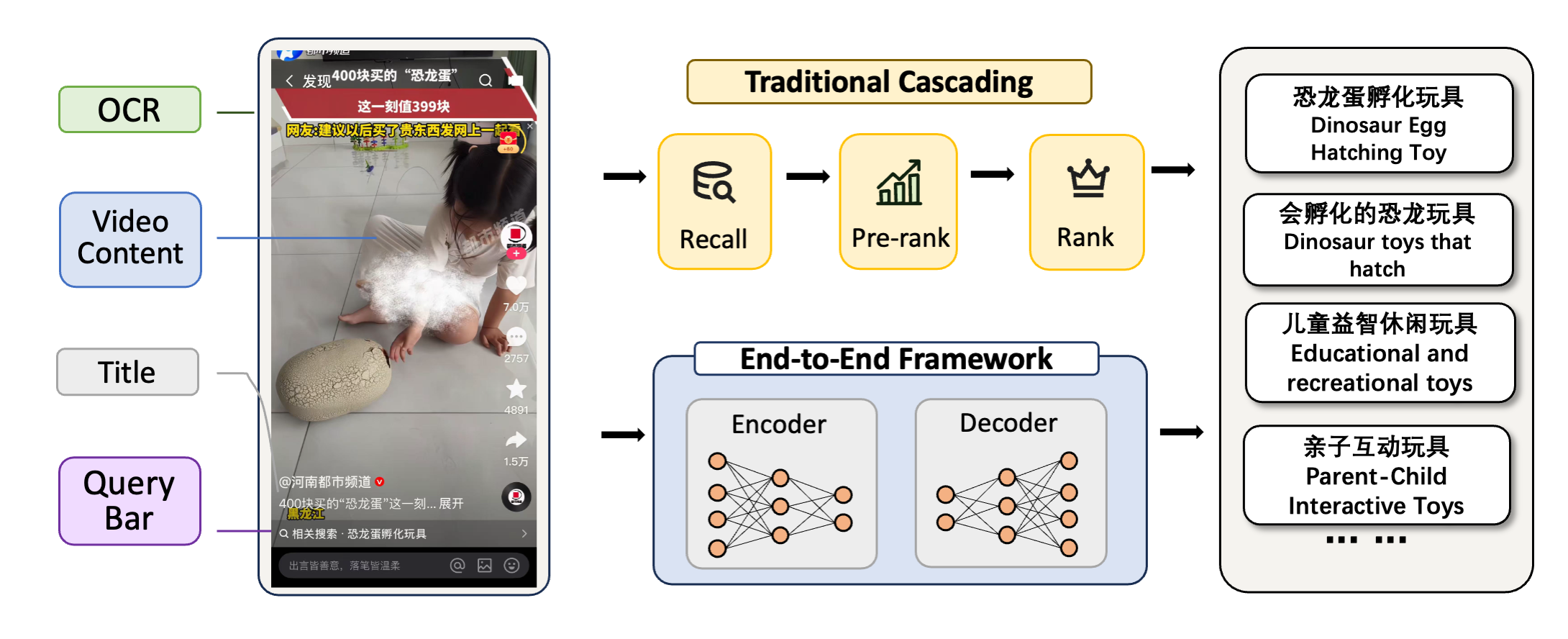}
    \caption{An example of query recommendation at bottom bar on user-viewing video, where our proposed end-to-end framework replaces the traditional cascade architecture to improve downstream conversion.}
    \label{fig:illustration}
\end{figure}

As short-video platforms attract massive user traffic, content-grounded query recommendation during video viewing has created opportunities for downstream advertising exposure and e-commerce transactions. For example, when a user is watching a makeup tutorial, a relevant query such as ``matte lipstick for dry lips'' can guide the user from video browsing to item search. As illustrated in Fig.~\ref{fig:illustration}, such recommendations mainly appear as an independent search bar beneath the video interface for further one-click searches, termed Bottom Bar Recommendation. Therefore, providing precise and personalized queries is crucial for additional conversion revenue.

Traditional query recommendation methods~\cite{xu2026aigq, min2025prompting}, which are not constrained by any context, typically rely on historical search logs, user profiles, or global popularity signals to suggest context-agnostic queries. However, directly applying these methods to short-video feeds may produce high CTR but content-irrelevant queries to the current video, leading to intent drift and degraded user experience. In short-video feeds, real user search intent is often induced by the content currently being viewed, so that recommended queries must be grounded in current video information.

\begin{figure}[tb]
    \centering
    \includegraphics[width=0.95\linewidth]{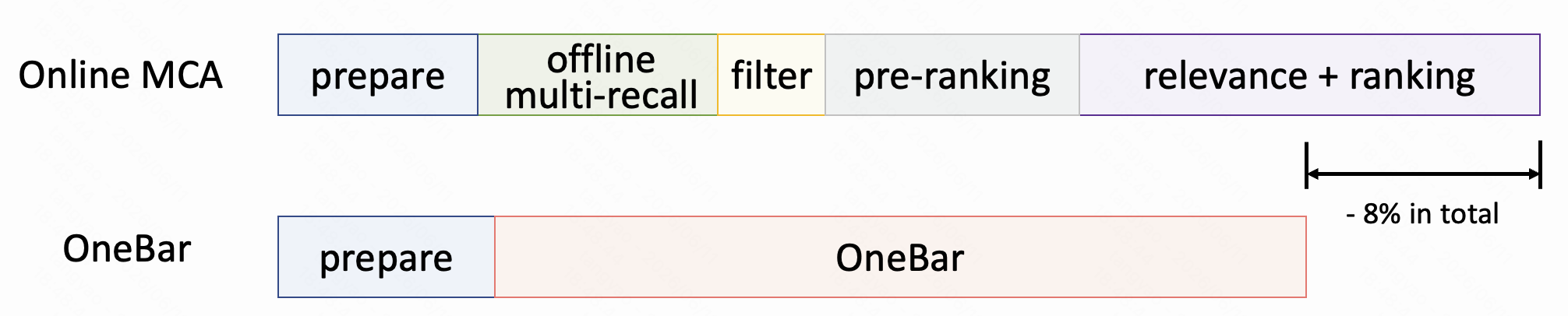}
    \caption{Online inference time comparison between online MCA and OneBar. OneBar keeps the same preparation stage and replaces the cascade stages with a single generation one.}
    \label{fig:time}
\end{figure}

Conventional industrial systems usually follow a retrieval-based multi-stage cascade architecture (MCA) ~\cite{bar2011context,huang2003relevant,sadikov2010clustering}. Candidate queries are first retrieved from historical query pools or behavior logs using lexical, semantic, or collaborative matching, and are then processed by downstream filtering, rewriting, and ranking modules. 
The bottom-bar scenario further poses stringent deployment challenges in industrial short-video feeds. Unlike standalone search services, the bottom-bar recommendation is displayed in real-time as users switch videos, which means that the system must handle extremely high QPS while satisfying a very tight serving latency budget. In the serving pipeline, query recommendation can only be executed after the short-video recommendation result is determined, but the recommended query must already be visible when the user starts watching the video. In our production system, this leaves only about 20--30 milliseconds for the bottom-bar service. Under such strict latency constraints, the online MCA pipeline can mainly rely on offline precomputed caching, so that enough time remains for ranking, scoring, relevance filtering, and risk-control filtering, as shown in Fig.~\ref{fig:time}. However, this design prevents the system from fully exploiting the real-time context of the user's current video-viewing session. As a result, the retrieved queries may lag behind newly uploaded videos, emerging trends, or rapidly changing user interests, and the offline candidate pool is often biased toward historically frequent queries. This makes it difficult to produce fresh, fine-grained, or long-tail queries that precisely match the current video content. Moreover, the multi-stage MCA pipeline fragments optimization across retrieval, filtering, relevance estimation, risk-control filtering, and ranking. While the final business objective depends on the joint effect of query relevance, exposure opportunity, CTR, and downstream conversion. Consequently, the online system often has to trade off query freshness and semantic relevance for serving efficiency, limiting its ability to provide high-quality bottom-bar recommendations at scale.

A natural way to overcome the above limitations is to replace the retrieval-based cascade with an end-to-end generative model, which can directly produce context-aware queries from the current video evidence. Recent efforts have begun to explore this direction~\cite{guo2026onesug}. However, in the short-video feed scenario, only a few closely related studies have been reported, and existing generative approaches still face important limitations. GREAT~\cite{shao2025great} improves query quality through trie-based constrained decoding, but its generation space remains bounded by the predefined query trie, which limits its ability to produce truly fresh queries. CLICKABLE~\cite{tianshort} introduces RAG-style behavioral context to incorporate historical signals, but its effectiveness depends heavily on the density and freshness of the retrieved evidence. Moreover, the latency introduced by multi-layer autoregressive decoding often restricts these methods to offline pre-computation or nearline serving, preventing them from fully leveraging real-time signals from the user's video-viewing context.

Beyond candidate expressiveness and serving efficiency, generative query recommendation in industrial short-video platforms is further complicated by two practical challenges. First, content-side signals are often noisy, incomplete, and sometimes misleading. Video titles,  @mentions, and promotional tags may contain redundant, promotional, or attention-seeking tokens that are only weakly related, or even irrelevant, to the actual video intent. As a result, directly conditioning on raw video-side metadata may mislead the generator and produce semantically drifted queries. Second, preference optimization becomes challenging when it relies on a separately trained reward model. Many generative recommendation pipelines~\cite{chen2025onesearch,deng2025onerec,xue2023prefrec} train an auxiliary reward model from historical user behavior logs to align generated queries with relevance, clicks, or conversions. However, such a reward model is inherently coupled with the logged data distribution and the training policy. It may therefore inherit biased feedback, amplify long-tail distributional biases, and be exploited through reward hacking~\cite{chen2026onesearch, liu2025onerec, zhou2025onerec}.

To address these limitations, we propose OneBar, an end-to-end generative framework for real-time query recommendation grounded in user-consumed short-video content. OneBar is driven by three key components.

\noindent\textbf{Collaborative-Multimodal Intent Grounding.}
OneBar abstracts each video impression into a structured evidence schema
$\mathcal{E}(x,u)=\langle T_x, M_x, A_x, H_{x,u}\rangle$,
which combines cleaned textual metadata, multimodal video summaries, collaborative query anchors and user history filtered by relevance. This module enables robust query generation under noisy short-video metadata.

\noindent\textbf{Low-Latency Unified End-to-End Generation.}
OneBar compresses heterogeneous evidence into a compact \texttt{[SEP]}-delimited schema and feeds it into a BART encoder--decoder generator. This unified formulation replaces fragmented multistage cascades and supports real-time online generation under strict latency constraints.

\noindent\textbf{Progressive Preference Internalization.}
OneBar first learns fine-grained intent preferences from hierarchical user interactions. It further introduces Preference-Internalized On-Policy Distillation (PIOPD) to inject posterior behavior signals directly into the policy, avoiding reliance on a separately trained reward model.

Extensive offline evaluations demonstrate OneBar's strong query relevance and search quality. We further conduct a rigorous online A/B test on Kuaishou's main video feed. For e-commerce videos, OneBar significantly improves key online metrics over online MCA, including +16.91\% Query Exposure, +18.68\% Query Click, +20.36\% guided orders, and +21.67\% guided GMV. Manual evaluation also shows that OneBar reduces the overall query-quality bad-case rate by 9.00 percentage points. To the best of our knowledge, OneBar is among the first end-to-end generative query recommendation systems deployed at scale for real-time online serving in integrated search-and-recommendation platforms.

In summary, this paper makes the following contributions:
\begin{itemize}

    \item We present OneBar, an end-to-end generative query recommendation framework for real-time bottom-bar serving in short-video feeds.

    \item We introduce Preference-Internalized On-Policy Distillation (PIOPD), which distills posterior preference signals directly into the generative policy and avoids reliance on a separately trained reward model.

    \item We validate OneBar through extensive offline evaluation and large-scale online A/B testing on Kuaishou Video Platform, demonstrating improvements in search quality, user engagement, search traffic, order volume, and GMV.

\end{itemize}

\begin{figure*}[t]
    \centering
    \includegraphics[width=\textwidth]{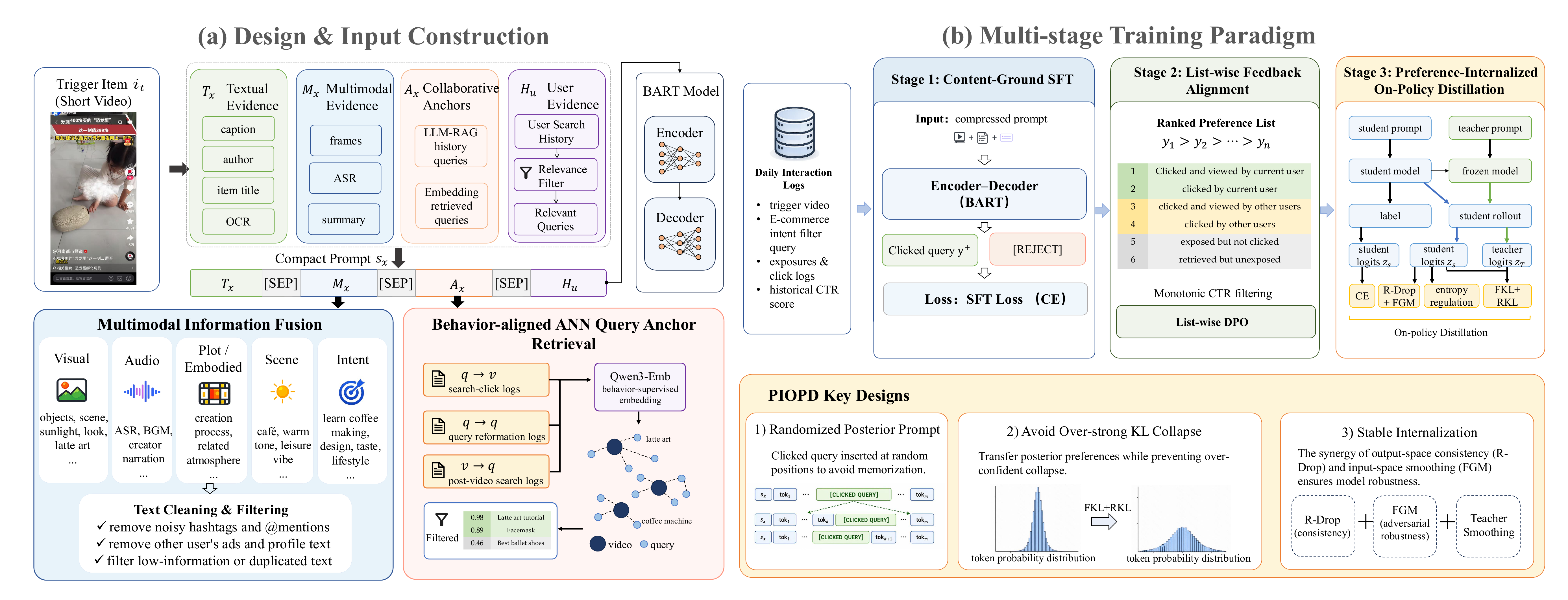}
    \caption{Overview of the OneBar framework.}
    \label{fig:framework}
\end{figure*}

\section{Related Work}
\label{relatedworks}

\noindent\textbf{Query Recommendation} and query suggestion have long been central tasks in search systems. Prior studies mainly differ in the type of context used to infer the user's search intent. A large body of work studies query auto-completion, where the system predicts complete queries from a typed prefix or partial input~\cite{bar2011context,shokouhi2013learning,li2014two,zhang2015adaqac,yuan2026unifying,wang2025personalized,guo2026onesug}. Another group of methods assumes that an initial query is already available and improves it through spelling correction~\cite{hagen2017large}, query expansion~\cite{carpineto2012survey}, or query rewriting~\cite{gottlob2011ontological,gottlob2014query,gong2026cardrewriter}. Beyond these query-driven settings, recent studies also explore proactive query recommendation from search logs, session context, user profiles, and behavior histories~\cite{baeza2004query,sordoni2015hierarchical,li2019click,bacciu2024generating,min2025prompting,xu2026aigq}. These methods are effective when user intent can be inferred from textual input or historical interactions, but they do not directly address the bottom-bar scenario in short-video feeds. In this setting, the recommendation is triggered by the currently consumed video rather than by an explicit search input, requiring the generated query to be grounded in multimodal content, aligned with video-induced intent, and served under strict latency constraints.

\noindent\textbf{Preference-Driven Policy Alignment} has become a central paradigm for adapting language models to task-specific objectives. Representative methods such as DPO~\cite{rafailov2023direct} learn directly from preference pairs without training an explicit reward model, while reinforcement-learning-based methods such as GRPO~\cite{shao2024deepseekmath} optimize policies through on-policy exploration. Recent studies further extend this idea to on-policy distillation, where supervision is applied to trajectories sampled from the current model~\cite{agarwal2024policy,zhao2026self,shenfeld2026self}. By aligning the training signal with the model's own generation distribution, on-policy distillation improves capability transfer and mitigates the exposure bias caused by purely teacher-forced training. Similar principles have also been explored in industrial search systems. OneSearch-V2~\cite{chen2026onesearch}  leverages online self-distillation to enhance the model's implicit reasoning and learning capabilities for generative search.

\section{Methodology}
\label{sec:methodology}

This section introduces OneBar, an end-to-end generative framework for real-time content-grounded query recommendation. We first formulate the bottom-bar query recommendation task in Section~\ref{sec:problem}. We then describe collaborative-multimodal intent grounding in Section~\ref{sec:context}, followed by the unified encoder--decoder generator in Section~\ref{sec:generator}. Section~\ref{sec:alignment} presents progressive preference internalization, which integrates content-grounded SFT, list-wise feedback alignment, and preference-internalized on-policy distillation. The overall framework is shown in Figure~\ref{fig:framework}.

\subsection{Problem Definition}
\label{sec:problem}

We study content-grounded query recommendation on a short-video platform. When a user watches a short video, the system proactively displays a small set of search queries in the bottom-bar search entry, guiding the user from passive content consumption to active search. For each impression, the system observes the current trigger item $x$ and optional user context $u$, and generates an ordered list of $K$ candidate queries:
\begin{equation}
Q_{x,u}=(q_1,q_2,\ldots,q_K),
\end{equation}
where each $q_i$ is a natural-language search expression. In our online deployment, $K=8$.

\begin{equation}
Q_{x,u}^{*}
=
\arg\max_{\substack{(q_1,\ldots,q_K)\in \Omega(x,u)^K\\ q_i\ne q_j,\ i\ne j}}
\sum_{i=1}^{K} w_i U(q_i;x,u),
\label{eq:utility-objective}
\end{equation}
where $U(q_i;x,u)$ denotes the posterior engagement utility of exposing query $q_i$ under the current trigger and user context, and $w_i$ is a position-dependent weight for the ordered bottom-bar list. The feasible set $\Omega(x,u)$ contains queries that satisfy production constraints:
\begin{equation}
\Omega(x,u)=
\{q \mid \mathrm{Ecom}(q),\ \mathrm{Rel}(q,x)\ge\delta,\ \mathrm{Safe}(q)\}.
\end{equation}

Here, $\mathrm{Ecom}(q)$ requires the query to express valid e-commerce search intent, $\mathrm{Rel}(q,x)$ measures its semantic relevance to the current short-video trigger, and $\mathrm{Safe}(q)$ requires the query to pass platform compliance checks, including GSB-based evaluation and prohibited-term filtering.

\subsection{Collaborative-Multimodal Intent Grounding}
\label{sec:context}

Before generation, OneBar converts each short-video trigger into a compact evidence representation. The representation combines textual metadata, multimodal content, collaborative query anchors, and user personalization signals, and is serialized into a concise schema consumed by generator.

\noindent\textbf{Textual and multimodal evidence ($T_x,M_x$).}
Raw short-video fields (captions, item titles, @mentions, hashtags) are noisy and often promotional. We form $T_x$ by selecting a canonical title from the first non-empty platform field (caption, photo title, title, or item title) and normalizing away user mentions, account identifiers, hashtags, and boilerplate. Because text alone misses content-grounded intent, we summarize sampled frames and ASR with a multimodal foundation model into a video summary $M_x$ that captures visual objects, OCR text, speech cues, scene context, product usage, and candidate search intent; $M_x$ is refreshed daily by an offline incremental pipeline and stripped of template-like prefixes.

\noindent\textbf{Collaborative Behavioral Evidence($A_x$).}
$A_x$ supplies behavior-derived intents that recur in similar video-consumption contexts and combines two separately stored fields. The first field consists of ANN-retrieved query anchors. To construct them, we adapt Qwen3-Embedding~\cite{zhang2025qwen3} 0.6B with staged behavior supervision over the three platform relations summarized in Table~\ref{tab:anchor-relations}. We first train the embedding model with the $q\!\rightarrow\!v$ relation from search-click logs, where a query leads to clicked and consumed videos, so as to build a relevance-preserving semantic space between queries and videos. Starting from this checkpoint, we further incorporate the $v\!\rightarrow\!q$ and $q\!\rightarrow\!q$ relations. The former captures queries issued after video consumption and therefore reflects actual search demand induced by the trigger, while the latter captures reformulation transitions in which follow-up queries often express clearer or more specific intent. The resulting model aligns videos and queries in a shared $128$-dimensional space whose proximity reflects observed query--video relevance. Offline, each trigger embedding retrieves the queries associated with its semantically nearest video triggers. Since low-similarity anchors may mislead the generator, we retain only anchors with trigger similarity $>\!0.88$, which raises the anchor hit rate from $0.282$ to $0.335$. The second field is RAG histories, which aggregate queries previously issued against the same trigger and serve as a complementary retrieval signal. Both fields are cached as generation evidence in $A_x$ rather than used as online candidate set.


\begin{table}[tb]
\centering
\caption{Behavioral relations for constructing collaborative query anchors $A_x$.}
\label{tab:anchor-relations}
\footnotesize
\setlength{\tabcolsep}{4pt}
\renewcommand{\arraystretch}{1.08}
\begin{tabularx}{\linewidth}{@{}l >{\raggedright\arraybackslash}X >{\raggedright\arraybackslash}X@{}}
\toprule
Relation & Source & Signal Type \\
\midrule
$q\!\rightarrow\!v$ & Search-click logs & Query $\rightarrow$ engaged video \\
$v\!\rightarrow\!q$ & Post-video search logs & Video $\rightarrow$ subsequent query \\
$q\!\rightarrow\!q$ & Query reformulation logs & Query $\rightarrow$ follow-up query \\
\bottomrule
\end{tabularx}
\end{table}

\noindent\textbf{Personalized evidence ($H_u$).}
Because bottom-bar intent is content-induced, indiscriminate global history weakens grounding. We therefore add only user-history queries whose similarity to the trigger, measured by the same behavior-aligned embedding model, exceeds a threshold, keeping $H_u$ tied to the trigger rather than to long-term interests.

\subsection{Low-Latency Unified End-to-End Generation with Abstention}
\label{sec:generator}
OneBar adopts a single encoder--decoder as its end-to-end generation backbone. The objective of this design is not to introduce a new architecture, but rather to provide a \emph{unified, low-latency, risk-controllable} interface that collapses the multi-stage retrieval cascade into a single end-to-end generation pass, thereby making real-time bottom-bar serving feasible within the 20--30\,ms budget specified in Sec.~\ref{sec:introduction}.
The four evidence fields---which are sparse in production, yet always contain at least one field to guarantee full coverage---are serialized into a single compact sequence delimited by \texttt{[SEP]}:
\begin{equation}
s_x=[\,T_x;\ \mathrm{[SEP]};\ M_x;\ \mathrm{[SEP]};\ A_x;\ \mathrm{[SEP]};\ H_u\,].
\label{eq:schema}
\end{equation}
This field-aligned schema is consistent with the text-reconstruction pretraining of BART and shortens the encoder input, thereby reducing serving latency; it substantially outperforms verbose instruction-style prompts, as reported in Table~\ref{tab:offline-ablation}. A BART backbone~\cite{lewis2020bart}, denoted $\mathcal{M}_\theta$, then defines the decoding distribution $p_\theta(y\mid s_x)=\prod_{t} p_\theta(y_t\mid y_{<t},\mathrm{Enc}_\theta(s_x))$.

The output $y$ is either a search query or the special token \texttt{[REJECT]}. The token \texttt{[REJECT]} allows the generator to \emph{abstain} when no safe, relevant, and confidently grounded query exists, e.g., under policy-sensitive content or weak trigger evidence. In this manner, the $\mathrm{Safe}(\cdot)$ predicate in Eq.~\eqref{eq:utility-objective} is enforced within the model itself, rather than by a downstream filter. Supervision for \texttt{[REJECT]} is introduced in Stage~I (Sec.~\ref{sec:alignment}); at serving time, a \texttt{[REJECT]} output suppresses bottom-bar exposure and accounts for a large portion of the reduction in online risk-control bad cases (Sec.~\ref{sec:eval}).
At inference time, OneBar decodes the bottom-bar list from $s_x$ using beam search with a width of $8$ under the latency budget, followed by lightweight de-duplication, safety, and business-rule checks before exposure.

\subsection{Progressive Preference Internalization}
\label{sec:alignment}

In content-grounded bottom-bar query recommendation, a generated query should not only be semantically relevant to the trigger, but also support platform-level utility, including click-through rate, downstream page consumption, and conversion. The generator must therefore remain consistent with the natural search distribution of the platform while capturing fine-grained preferences observed from online feedback. We address this problem through a progressive preference internalization procedure. Content-grounded SFT first establishes the basic mapping from triggers to queries, list-wise feedback alignment then learns relative preferences among candidates, and PIOPD finally transfers dense posterior preferences into the generator.

\subsubsection{Stage I: Content-Grounded SFT}

We first apply supervised fine-tuning (SFT) to initialize the generator with content-grounded query generation ability. This stage encourages the model to produce natural search expressions that are relevant to the current trigger and suitable for e-commerce search. We construct trigger-query pairs from online click logs, where the clicked query is treated as the positive target for the corresponding trigger. To keep the training data focused on e-commerce intent, we further apply a platform-side e-commerce intent classifier and remove queries that do not express valid item-search intent. The retained samples form
$\mathcal{D}_{\mathrm{pos}}=\{(s_x,y^+)\}$, where $s_x$ denotes the serialized trigger evidence and $y^+$ denotes the clicked query.

Direct training on online behavior data may introduce safety risks, because user clicks can occasionally favor sensational, inappropriate, or policy-violating expressions. Although such queries may receive high engagement, displaying them in the bottom-bar entry would degrade user experience and weaken user trust. We therefore introduce safety-aware rejection supervision. We collect video impressions whose content or associated textual fields contain non-compliant signals identified by platform-side safety rules, including prohibited-term matching and GSB-based evaluation. For these samples, the target output is set to the special token \texttt{[REJECT]} instead of a normal query, forming
$\mathcal{D}_{\mathrm{rej}}=\{(s_x,\texttt{[REJECT]})\}$.

The generator is trained with standard next-token prediction:

\begin{equation}
\begin{aligned}
\mathcal{L}_{\mathrm{SFT}}
&= -\mathbb{E}_{(s_x,y^+)}
\sum_{t=1}^{|y^+|} 
\log p_\theta(y_t^+\mid y_{<t}^+,s_x).
\end{aligned}
\end{equation}

During online serving, if the generator outputs \texttt{[REJECT]}, the system suppresses bottom-bar query exposure for the current impression. This mechanism allows the model to learn both what to recommend and when to abstain from recommendation.

\subsubsection{Stage II: Behavior Feedback Preference Alignment}

Online serving logs provide large-scale implicit feedback on query candidates generated for real user requests. Rather than converting such feedback into independent binary labels, we organize candidates with the same trigger into a behavior-induced ordinal preference list and optimize the generator with a list-wise objective. This enables learning from real online preferences while preserving the relative structure of implicit feedback.

For each trigger $x$, we collect its associated query candidates from online logs and assign them to six behavior levels, as summarized in Table~\ref{tab:behavior_levels}. The levels encode two preference principles. First, feedback from the current user provides stronger evidence of personalized intent than feedback aggregated from other users. Second, post-click engagement indicates stronger intent than clicks alone. Therefore, candidates clicked and further engaged by the current user are placed at the highest level, followed by candidates clicked by the current user, crowd-level engaged clicks, and crowd-level clicks. Exposed-but-unclicked candidates and retrieved-but-unexposed candidates are placed at lower levels and are used as contrastive alternatives.

To reduce noise introduced by online ranking, exposure position, and stochastic user interactions, we further enforce consistency between the behavior order and historical CTR. Candidates whose CTR scores contradict the monotonic order implied by their behavior levels are filtered out, since such conflicts usually indicate unreliable or weak preference evidence. After de-duplication, if multiple candidates remain in the same level, we keep the one with the highest historical CTR as the representative of that level, so that each level contributes a single, behavior-consistent preference target. For each trigger $x$, the retained candidates form an ordered preference list
$\mathcal{Y}_x=[y_1,\ldots,y_n]$, where $y_1 \succ y_2 \succ \cdots \succ y_n$ and $n\le 6$. This list provides ordinal supervision over candidate queries rather than calibrated utility scores.

\begin{table}[tb]
\centering
\caption{Hierarchical user behavior levels.}
\label{tab:behavior_levels}
\begin{tabularx}{\linewidth}{>{\centering\arraybackslash}p{0.12\linewidth}X}
\toprule
\textbf{Level} & \textbf{Behavior Description} \\
\midrule
1 & Clicks with post-click engagement by the current user \\
2 & Clicks by the current user \\
3 & Clicks with post-click engagement by other users \\
4 & Clicks by other users \\
5 & Exposed but unclicked \\
6 & Retrieved but unexposed \\
\bottomrule
\end{tabularx}
\end{table}

Given the behavior-induced preference list, we adopt a list-wise Softmax DPO objective~\cite{chen2024softmax} to align the generator with online user feedback. Let
$\ell_\theta(y\mid x)=\sum_t \log \pi_\theta(y_t\mid x,y_{<t})$
denote the sequence log-probability under the current policy, and let
\begin{equation}
r_i=\ell_\theta(y_i\mid x)-\ell_{\mathrm{ref}}(y_i\mid x)
\end{equation}
be the policy-reference log-ratio for candidate $y_i$. Let $b(y_i)$ denote the behavior level of $y_i$, and define the clicked anchors as
\begin{equation}
\mathcal{C}_x=\{j \mid b(y_j)\le 4,\ j<n\}.
\end{equation}
Only clicked candidates are used as positive anchors, while lower-ranked candidates, including non-clicked ones, serve as contrastive alternatives.

For each clicked anchor $y_j$ with $j\in\mathcal{C}_x$, we compare it against all lower-ranked candidates in the same trigger-level list, denoted by
$\mathcal{R}_j=\{y_{j+1},\ldots,y_n\}$. Under a softmax choice model over $\{y_j\}\cup\mathcal{R}_j$, the probability that $y_j$ is preferred to its lower-ranked alternatives is
\begin{equation}
P_\theta(y_j \succ \mathcal{R}_j\mid x)
=
\frac{\exp(\beta r_j/T)}
{\exp(\beta r_j/T)+\sum_{i=j+1}^{n}\exp(\beta r_i/T)}.
\end{equation}
The corresponding negative log-likelihood is
\begin{equation}
-\log P_\theta(y_j \succ \mathcal{R}_j\mid x)
=
\log\left(1+\sum_{i=j+1}^{n}
\exp\left(\frac{\beta}{T}(r_i-r_j)\right)\right)
\end{equation}

Compared with pair-wise DPO, which constructs isolated chosen--rejected pairs, this objective uses the full lower-ranked suffix as the contrastive context for each clicked candidate. It therefore transfers multi-level online behavior preferences into the generator while avoiding positive supervision from non-clicked candidates.

The Stage-II training objective is defined as
\begin{equation}
\begin{split}
\mathcal{L}_{\mathrm{Stage\text{-}II}}(x)
= \sum_{j\in\mathcal{C}_x} \Bigg[
& \log\left(1 + \sum_{i=j+1}^{n}
\exp\left( \frac{\beta}{T}(r_i-r_j) \right) \right) \\
& - \lambda_{\mathrm{SFT}} \ell_\theta(y_j\mid x)
\Bigg].
\end{split}
\end{equation}
Here, $\beta=0.1$ is the DPO scale factor, $T=1.0$ is the softmax temperature, and $\lambda_{\mathrm{SFT}}=1.0$ controls the SFT anchor. The SFT term is applied to clicked anchors to stabilize preference optimization and preserve the generator's sequence modeling ability. In implementation, candidates from the same trigger are packed into a single forward pass, enabling efficient list-wise preference learning from large-scale online behavior logs.

\subsubsection{Stage III: Preference-Internalized On-Policy Distillation}

Stage-II preference alignment improves the generator through behavior-derived sequence-level supervision, but the generator is still optimized on logged candidates. This may leave an off-policy gap between the training data and the inference-time distribution of the model. To reduce this gap, we introduce \textit{Preference-Internalized On-Policy Distillation} (PIOPD), which performs dense token-level distillation on trajectories sampled from the current student policy.

PIOPD uses posterior behavior evidence as training-time privileged supervision in order to internalize user preference into the deployable generator without changing its online interface. Unlike OneSearch-V2~\cite{chen2026onesearch}, which augments the teacher input with Chain-of-Thought reasoning paths, we use clicked target queries as posterior preference signals. During training, the teacher receives the same video information as the student, together with the clicked query associated with the video. This posterior query identifies a behavior-confirmed intent region and enables the teacher to assign soft probability mass to the tokens and query variants that are consistent with the observed user preference. The student is then optimized to match the soft token distributions of the teacher on student-sampled prefixes, although it never observes the clicked query itself. As a result, the posterior evidence influences the final model only through parameter learning: PIOPD does not require an additional reward model for online scoring, nor does it introduce a posterior-conditioned module into the serving pipeline. The deployed artifact remains a single generator conditioned only on the standard trigger and user context.

\begin{figure}[tb]
\centering
\includegraphics[width=0.95\linewidth]{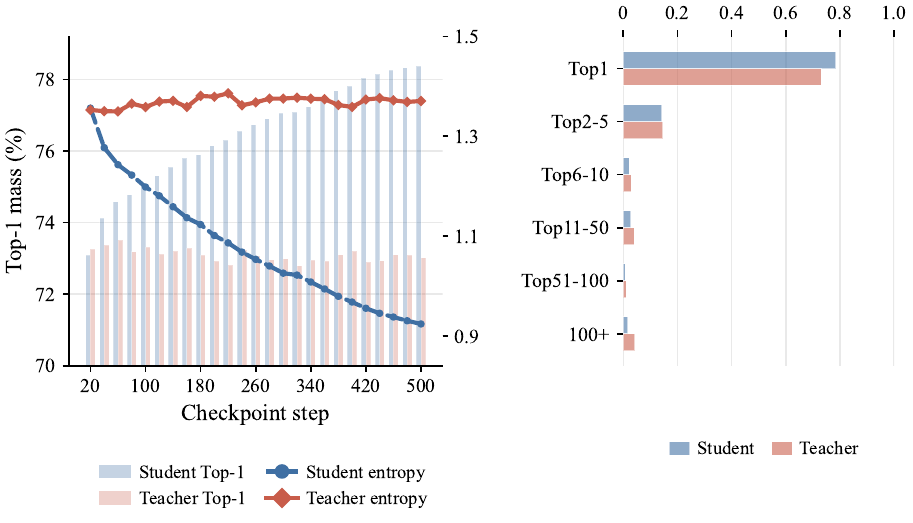}
\caption{Token-level distribution analysis.}
\label{fig:entropy-top1-prob-distribution}
\end{figure}
We build the posterior-aware teacher by augmenting its input with the clicked query. Appending $y_{\text{ref}}$ at a fixed position can push the teacher toward positional shortcuts or surface-form copying, and we therefore adopt Randomized Context Augmentation, which inserts the clicked query at a randomly sampled position. This encourages the teacher to treat $y_{\text{ref}}$ as an intent-level posterior signal rather than a fixed suffix. Let $\bm{x}$ denote the standard context; the teacher receives $\bm{x}^{(T)}$ and the student receives $\bm{x}^{(S)}$:
\begin{equation}
\bm{x}^{(T)} = \bm{x}\oplus_{\text{rand}} y_{\text{ref}},
\qquad
\bm{x}^{(S)} = \bm{x}.
\end{equation}
As reported in Table~\ref{tab:teacher-prompt}, randomized insertion yields the highest teacher HR among the tested prompt variants, suggesting that distributing the posterior evidence across the input provides a more effective signal than fixed-position injection. We use this teacher performance only to validate the construction of the teacher; all deployable results are reported for the student under the standard prompt.

We project the posterior-augmented teacher onto the deployable student policy over student-visited states:

\begin{equation}
\begin{aligned}
\theta^*
= \arg\min_{\theta}\ 
\mathbb{E}_{y\sim \pi_\theta(\cdot\mid x^{(S)})}
\sum_t
D\Bigl(
&\pi_T^+(\cdot\mid y_{<t},x^{(T)}) \\
&\big\|\
\pi_\theta(\cdot\mid y_{<t},x^{(S)})
\Bigr).
\end{aligned}
\label{eq:piopd-projection}
\end{equation}

Since the expectation is taken over student rollouts rather than over teacher-forced gold prefixes, the projection is performed on the states that the deployed generator is likely to visit at inference time. In practice, we sample one trajectory autoregressively from the current student under $\bm{x}^{(S)}$. At each sampled prefix $y_{<t}$, the student and the frozen teacher produce logits $\bm{z}_t^{(S)}$ and $\bm{z}_t^{(T)}$ over the vocabulary $\mathcal{V}$, with the teacher conditioned on $\bm{x}^{(T)}$ and the student conditioned on $\bm{x}^{(S)}$. Let $p_S^t=\mathrm{softmax}(\bm{z}_t^{(S)})$ and $p_{T,\tau}^t=\mathrm{softmax}(\bm{z}_t^{(T)}/\tau)$ with $\tau=2$. We instantiate $D(\cdot\Vert\cdot)$ in Eq.~\eqref{eq:piopd-projection} with a bidirectional KL objective:
\begin{equation}
\begin{aligned}
\mathcal{L}_{\mathrm{Distill}}
= \mathbb{E}_{y \sim \pi_\theta(\cdot \mid \bm{x}^{(S)})}
\sum_t \tau^2 \Big[
& \lambda_{\mathrm{FKL}}
D_{\mathrm{KL}}(p_{T,\tau}^t \Vert p_S^t) \\
+\,& \lambda_{\mathrm{RKL}}
D_{\mathrm{KL}}(p_S^t \Vert p_{T,\tau}^t)
\Big].
\end{aligned}
\end{equation}
The two terms play complementary roles. The forward KL term encourages the student to cover the alternatives supported by the teacher and penalizes missing mass on high-probability teacher tokens, whereas the reverse KL term suppresses student mass on teacher-unsupported tokens and reduces mode drift. We retain the SFT loss $\mathcal{L}_{\mathrm{SFT}}$ as a ground-truth anchor, so that the student absorbs posterior preferences without losing its base generation ability. Table~\ref{tab:offline-ablation} shows that combining FKL, RKL, and SFT outperforms any single component.

A common failure mode in posterior distillation is the gradual over-concentration of the student distribution. As Fig.~\ref{fig:entropy-top1-prob-distribution} shows, the teacher remains calibrated across checkpoints, whereas the student shifts mass toward the top-1 token: from step~20 to step~400, the top-1 mass rises by 5.0\%, and the tail mass beyond top-10 drops by 4.4\%. This over-amplification reduces diversity and harms correct alternatives that are not ranked first. We therefore add an entropy regularizer on the student:
\begin{equation}
\begin{aligned}
\mathcal{L}_{\mathrm{Ent}}
&= -\mathbb{E}_{y\sim \pi_\theta(\cdot \mid \bm{x}^{(S)})}
\sum_t H(p_S^t) \\
&= \mathbb{E}_{y \sim \pi_\theta(\cdot \mid \bm{x}^{(S)})}
\sum_t \sum_{v \in \mathcal{V}}
p_S^t(v) \log p_S^t(v).
\end{aligned}
\end{equation}
Since $\mathcal{L}_{\mathrm{Ent}}$ is the negative entropy, minimizing it with a positive weight $\lambda_{\mathrm{E}}$ encourages higher student entropy and preserves generation diversity. Guided by the business need for stable high-recall candidate generation under noisy behavior feedback, we further apply output-space and input-space stabilization through R-Drop~\cite{wu2021r} and the Fast Gradient Method (FGM)~\cite{miyato2016adversarial}, respectively. R-Drop improves output-space consistency across two student passes with independent dropout masks, and FGM provides input-space smoothing through an adversarial embedding perturbation $\Delta \bm{e} = \epsilon \cdot \nabla_{\bm{e}} \mathcal{L}_{\mathrm{task}} / \Vert \nabla_{\bm{e}} \mathcal{L}_{\mathrm{task}} \Vert$:
\begin{equation}
\begin{aligned}
\mathcal{L}_{\mathrm{RDrop}}
&= \tfrac{1}{2} \big[
D_{\mathrm{KL}}(p_{S_1}\Vert p_{S_2})
+ D_{\mathrm{KL}}(p_{S_2} \Vert p_{S_1})
\big], \\
\mathcal{L}_{\mathrm{FGM}}
&= \mathcal{L}_{\mathrm{task}}(\bm{e} + \Delta\bm{e}).
\end{aligned}
\end{equation}

The final objective combines all components,
\begin{equation}
\begin{aligned}
\mathcal{L}_{\mathrm{PIOPD}}
&= \mathcal{L}_{\mathrm{SFT}} + \mathcal{L}_{\mathrm{Distill}}
+ \lambda_{\mathrm{E}}\mathcal{L}_{\mathrm{Ent}} \\
&\quad
+ \lambda_{\mathrm{RD}}\mathcal{L}_{\mathrm{RDrop}}
+ \lambda_{\mathrm{FGM}}\mathcal{L}_{\mathrm{FGM}},
\end{aligned}
\end{equation}
and the full procedure is summarized in Algorithm~\ref{alg:piopd}. This objective allows the student to internalize posterior preferences while maintaining a diverse generation distribution. After training, we discard the teacher and the posterior signal $y_{\text{ref}}$ and retain only the deployment-consistent student, so that the serving interface of Sec.~\ref{sec:generator} remains unchanged.
\begin{algorithm}[tb]
\caption{Preference-Internalized On-Policy Distillation (PIOPD)}
\label{alg:piopd}
\begin{algorithmic}[1]
\REQUIRE Training instances $(\bm{x}, y_{\mathrm{ref}})$, where $\bm{x}$ is the standard trigger context and $y_{\mathrm{ref}}$ is the clicked target query; Stage-II student $\pi_\theta$
\ENSURE Deployable student $\pi_\theta$ conditioned only on $\bm{x}$
\STATE Initialize $\pi_\theta$ from the Stage-II checkpoint
\STATE Initialize frozen teacher $\pi_T^+$ from Stage-II checkpoint
\WHILE{not converged}
\STATE Sample a minibatch ${(\bm{x}, y_{\mathrm{ref}})}$
\STATE Set student input $\bm{x}^{(S)}=\bm{x}$
\STATE Set teacher input $\bm{x}^{(T)}=[\bm{x}; y_{\mathrm{ref}}]$
\STATE Generate an on-policy rollout $y^{\mathrm{roll}}\sim\pi_\theta(\cdot\mid \bm{x}^{(S)})$
\STATE Compute student rollout logits
$\bm{z}^{(S,r)} \leftarrow \pi_\theta(\cdot\mid y^{\mathrm{roll}}, \bm{x}^{(S)})$
\STATE Compute teacher rollout logits
$\bm{z}^{(T,r)} \leftarrow \pi_T^{+}(\cdot\mid y^{\mathrm{roll}}, \bm{x}^{(T)})$
\STATE Compute student label logits
$\bm{z}^{(S,l)} \leftarrow \pi_\theta(\cdot\mid y_{\mathrm{ref}}, \bm{x}^{(S)})$
\STATE Compute $\mathcal{L}_{\mathrm{PIOPD}}$ from
$\bm{z}^{(S,r)}$, $\bm{z}^{(T,r)}$, and $\bm{z}^{(S,l)}$
\STATE Update $\theta$ by gradient descent on $\mathcal{L}_{\mathrm{PIOPD}}$
\ENDWHILE
\STATE Discard $\pi_T^{+}$ and the posterior signal $y_{\mathrm{ref}}$
\end{algorithmic}
\end{algorithm}

\section{Online Deployment}
\label{sec:online}

We deploy OneBar in a hybrid offline-online serving architecture. The design decouples expensive evidence construction from latency-critical query generation.

For each video trigger, we cache pre-computed evidence in a Redis-backed feature store keyed by \texttt{photo\_id}. The cached evidence includes video embeddings produced by the dual-tower retrieval model, daily-refreshed RAG histories, and multimodal features extracted from video content. This avoids invoking expensive retrieval, LLM, or multimodal components in online serving path.

At online inference time, OneBar fetches cached video evidence and assembles collaborative and personalized signals using offline-maintained embeddings and retrieval artifacts. This avoids expensive representation construction in the latency-critical path while preserving behavior-derived query anchors and trigger-related user history. The assembled evidence is fed into the BART-based encoder--decoder generator for synchronous real-time serving. We use beam search with a beam width of 8, followed by safety filtering, and business-rule checks before exposure.

To track content and feedback shifts, we daily update the model with newly collected online logs through supervised fine-tuning and PIOPD-based preference distillation, followed by evaluation and online refresh.

\section{Evaluation}
\label{sec:eval}

\begin{table*}[!t]
\centering
\begingroup
\footnotesize
\setlength{\tabcolsep}{4pt}
\renewcommand{\arraystretch}{1.08}
\caption{Offline generation-quality metrics. ED-HR@8 counts a query as a hit when the edit distance to a positive query is no larger than 2. Lower ED@8 is better.}
\label{tab:supplementary-metrics}
\begin{tabularx}{\textwidth}{X c c c c c c c c}
\toprule
Variant & LR & Exact HR@8 & MRR@8 & ED-HR@8 & ED@8 $\downarrow$ & BLEU@8 & BAS@8 & SR \\
\midrule
GLM5.1 (zero-shot) & -- & 0.0153 & 0.0036 & 0.0837 & 6.7051 & 0.1317 & 0.6468 & 4.4694 \\
GPT-5.5 (zero-shot) & -- & 0.0224 & 0.0139 & 0.1384 & 5.2667 & 0.1779 & 0.6261 & 4.9286 \\
\midrule
ANN & -- & 0.1322 & 0.0739 & 0.3052 & 4.0630 & 0.3218 & 0.8326 & 2.4759 \\
BART + Basic Video Information & $1{\times}10^{-5}$ & 0.1787 & 0.0805 & 0.2490 & 5.2471 & 0.2532 & 0.5839 & 2.8537 \\
\midrule
Stage 1: Context-Grounded SFT & $1{\times}10^{-5}$ & 0.3564 & 0.2398 & 0.4864 & 3.1785 & 0.4951 & 0.6678 & 3.8019 \\
Stage 2: List-wise Feedback Alignment & $5{\times}10^{-5}$ & 0.3586 & 0.1789 & 0.4470 & 3.5556 & 0.4489 & 0.6649 & 3.7488 \\
Stage 3: Progressive Preference Internalization & $1{\times}10^{-6}$ & \textbf{0.3690} & \textbf{0.2402} & \textbf{0.4934} & \textbf{3.1760} & \textbf{0.5039} & \textbf{0.6687} & 3.7945 \\
\bottomrule
\end{tabularx}
\endgroup
\end{table*}

To assess the OneBar, we conduct offline evaluations and online A/B tests in a real-world industrial setting, complemented by ablation studies and diagnostic analyses on its major design components.

\subsection{Experimental Settings}

\noindent\textbf{Dataset.}
We build an offline dataset from production logs collected between Apr.~9 and Apr.~16, 2026, comprising approximately 40 million page views with bottom-bar query exposures; the first seven days are used for training. To prevent information leakage from inflating offline metrics, we decontaminate the evaluation inputs by removing any historical inputs (e.g., RAG histories or embedding-retrieved anchors) that explicitly contain the target query.

\begin{table}[!t]
\centering
\caption{Comprehensive offline ablation study.}
\label{tab:offline-ablation}
\begin{tabularx}{\linewidth}{X c c}
\toprule
\textbf{Variant} & \textbf{HR@8} & \textbf{MRR@8} \\
\midrule
\multicolumn{3}{l}{\textit{Block 1: Feature Augmentation}} \\
BART + Basic Trigger Information & 0.1787 & 0.0805 \\
\quad + Multimodal Evidence ($M_x$) & 0.3046 & 0.1694 \\
\quad + Collaborative Anchors ($A_x$) & 0.3357 & 0.2098 \\
\quad + Personalized Evidence ($H_u$) & 0.3564 & 0.2398 \\
\midrule
\multicolumn{3}{l}{\textit{Block 2: Prompt Format Engineering}} \\
BART + Verbose Prompt (SFT) & 0.1864 & 0.1059 \\
BART + Compact \texttt{[SEP]} Schema & 0.3564 & 0.2398 \\
\midrule
\multicolumn{3}{l}{\textit{Block 3: Progressive Preference Internalization}} \\
Stage II: List-wise Feedback Alignment & 0.3586 & 0.1789 \\
Stage III: Base (0.5 RKL + 0.5 FKL) & 0.3630 & 0.2399 \\
0.5 RKL + 0.5 FKL + SFT & 0.3654 & \textbf{0.2408} \\
0.7 RKL + 0.5 FKL + SFT & 0.3669 & 0.2403 \\
\quad + Teacher Smoothing ($\tau{=}2$) & 0.3673 & 0.2399 \\
\quad + R-Drop + FGM ($\epsilon{=}0.6$) & 0.3675 & 0.2404 \\
Full PIOPD (+ Entropy Regularization, $\lambda_E{=}0.3$) & \textbf{0.3690} & 0.2402 \\
\bottomrule
\end{tabularx}
\end{table}

\noindent\textbf{Baselines.}
We establish three categories of baselines. (i) A \textit{Generative Base Model} serves as our primary generative baseline: a vanilla model trained with supervised fine-tuning (SFT) on basic trigger information only, without any feature augmentation. (ii) \textit{Zero-shot LLMs}, where we select state-of-the-art models spanning both open-source and proprietary ecosystems, including GLM5.1 and GPT-5.5, each prompted to generate recommendations directly without domain adaptation or fine-tuning. (iii) An \textit{ANN retrieval} baseline, which encodes the trigger evidence with the same collaborative-data-trained embedding model and retrieves the top-$K$ most similar queries from the historical query corpus via approximate nearest neighbor search. This provides a non-generative, candidate-retrieval reference that shares OneBar's evidence representation but lacks generative composition, isolating the contribution of generation over pure retrieval.

\noindent\textbf{Evaluation Metrics.}
Since end-to-end query generation should be evaluated beyond exact retrieval-style matching, we assess generated queries from complementary perspectives.
For \textit{ranking and recall}, we report Hit Rate@$K$ (HR@$K$) and Mean Reciprocal Rank (MRR) based on exact matches with the ground-truth query.
For \textit{intent accuracy}, since semantically equivalent queries may take different surface forms, exact-match HR can underestimate generation quality; we therefore add Edit-Distance-tolerant Hit Rate@$K$ (ED-HR@$K$), which counts a prediction as correct when its Levenshtein distance to the target is no larger than 2, and sentence-level BLEU@$K$~\cite{papineni2002bleu}, taken as the best BLEU among the top-$K$ generated queries.
For \textit{relevance}, we report two complementary metrics. Behavior-Aligned Similarity@$K$ (BAS@$K$) is measured by a collaborative-data-trained embedding model: we compute the similarity between the trigger evidence and each of the top-$K$ generated queries and report the average. Semantic Relevance (SR) is rated by GPT-5.5 as an LLM judge on a 0--5 scale, assessing whether a query is searchable and whether its intent is semantically related to the trigger video.

\noindent\textbf{Implementation Details.}
To balance query utility and candidate diversity, we use beam search decoding for both offline evaluation and online serving. The model is trained progressively in three stages. In Stage~I, the generator is supervised fine-tuned for 6 epochs with a batch size of 128 and a learning rate of $1\times10^{-5}$. In Stage~II, we apply behavior-feedback preference alignment with a learning rate of $5\times10^{-5}$ and a DPO scale factor of $\beta=0.1$. In Stage~III, we conduct PIOPD for 500 steps with a batch size of 64 and a learning rate of $1\times10^{-6}$. The trainable student and the frozen teacher share the same architecture and are both initialized from the Stage~II checkpoint, with the teacher distribution softened via temperature smoothing ($\tau=2$).

\subsection{Offline Evaluation}

\noindent\textbf{Overall Offline Performance.}
Table~\ref{tab:supplementary-metrics} shows that the full OneBar model (Stage 3) achieves the best results on the core matching-oriented metrics, including Exact HR@8, MRR@8, ED-HR@8, ED@8, and BLEU@8. The consistent gains over zero-shot LLMs, ANN retrieval, and basic BART indicate that both domain adaptation and trigger-grounded evidence are necessary for industrial bottom-bar query generation.

BAS@$K$ and SR require separate interpretation. ANN obtains the highest BAS@$K$ because the metric is computed with the same collaborative embedding model used by ANN retrieval, which introduces a self-consistency advantage. Excluding this structural bias, OneBar achieves the best BAS@$K$ among learning-based variants. For SR, zero-shot LLMs score highest under the LLM judge due to their fluent and topically coherent outputs, but their weak Exact HR@8, ED-HR@8, and BLEU@8 show that such outputs rarely match real user queries. In contrast, OneBar maintains competitive SR while substantially improving behavior-grounded accuracy.

Across training stages, Stage 3 improves Exact HR@8, MRR@8, ED-HR@8, ED@8, and BLEU@8, although Stage 1 has a slightly higher SR. This reflects the intended effect of PIOPD: it trades a small amount of single-output semantic sharpness for stronger candidate-set coverage and matching accuracy.

\noindent\textbf{Ablation Study.}
To examine the effectiveness of the major design choices in the OneBar framework, we conduct ablation studies using exact HR@8 and MRR@8 as main offline metrics. As shown in Table~\ref{tab:offline-ablation}, the ablation is structured into three blocks: feature augmentation, prompt format engineering, and progressive preference internalization.

The first block shows that each additional evidence source contributes to content-grounded query generation. Compared with the basic-trigger baseline, adding multimodal evidence, collaborative anchors, and personalized evidence consistently improves offline performance, confirming that OneBar benefits from complementary trigger signals rather than a single metadata field. These gains are complementary: multimodal evidence mitigates noisy or incomplete metadata, collaborative anchors add behavior-derived intent priors, and personalized evidence narrows generation toward trigger-relevant user interests.

The prompt-format block shows that, under the same Content-Grounded SFT recipe, the compact \texttt{[SEP]} schema substantially outperforms the verbose prompt (0.3564 vs. 0.1864 HR@8), indicating that field-aligned evidence is easier for BART to exploit than instruction-like wrappers.

The third block studies how behavior preferences are progressively absorbed by the generator. Stage II introduces list-wise feedback alignment and slightly improves HR@8 over Stage I: Content-Grounded SFT. Applying PIOPD directly after Content-Grounded SFT improves HR@8 from 0.3564 to 0.3628, while the full Content-Grounded SFT $\rightarrow$ Stage II $\rightarrow$ PIOPD pipeline further increases HR@8 to 0.3690, which is 0.0062 higher than skipping Stage II. In contrast, Stage II alone substantially reduces MRR@8 from 0.2398 to 0.1789, indicating that list-wise behavior alignment can move generated candidates toward behavior-preferred regions but does not preserve fine-grained ranking quality when used alone. After PIOPD is applied on top of Stage II, MRR@8 recovers to 0.2402 and HR@8 reaches the best value of 0.3690. These results suggest that Stage II mainly reshapes the student rollout distribution into a more behavior-aligned on-policy region, thereby providing a better starting point for token-level posterior distillation rather than serving as the final aligned generator.

\begin{figure}[!t]
\centering
\includegraphics[width=0.85\linewidth]{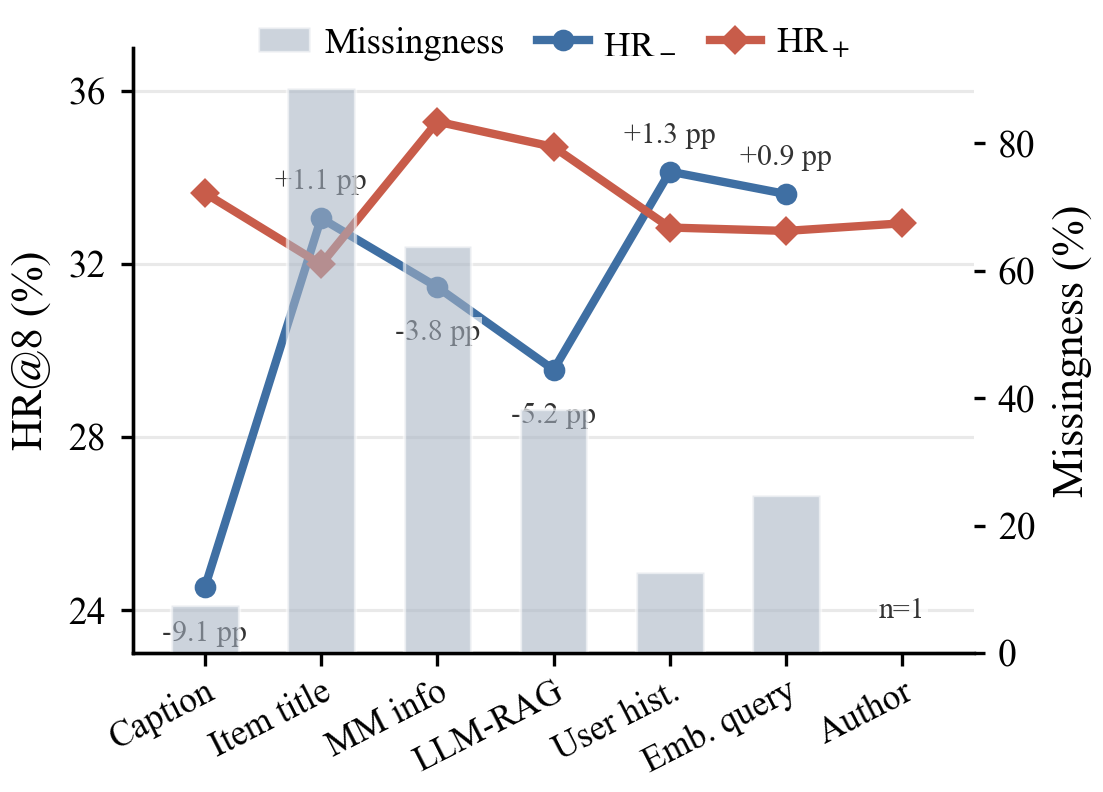}
\caption{Evidence-field missingness and HR@8 under availability slices. Gray bars show empirical missingness, while HR$_-$ and HR$_+$ denote HR@8 when the field is absent or present, respectively.}
\label{fig:field-missing-hr}
\end{figure}

Figure~\ref{fig:field-missing-hr} evaluates evidence fields under natural production missingness rather than controlled removal. Caption availability gives the clearest textual gain: the HR@8 of the caption-absent slice is about 24.5, while the caption-present slice reaches about 33.6. Multimodal summaries also provide a strong signal, with HR@8 rising from about 31.5 when absent to about 35.3 when present. For the collaborative part of $A_x$, RAG histories are especially important in behavior-sparse cases: the slice without RAG histories stays around 29.5 HR@8, while the slice with RAG histories reaches about 34.7. Embedding-query anchors and filtered user history have smaller aggregate effects because their availability is biased toward harder behavior-rich or personalized cases; their present-slice HR@8 is about 32.7 and 32.8, respectively, rather than uniformly above the absent slices. This slice-level behavior explains why the full-data increments of $A_x$ and $H_u$ are modest in Table~\ref{tab:offline-ablation}: they act as auxiliary priors for difficult or sparse contexts instead of standalone shortcuts. The fixed \texttt{[SEP]} schema keeps this robustness compatible with serving because missing fields do not change the input format.

\begin{table}[tb]
\centering
\caption{Teacher-prompt candidates used for PIOPD teacher selection. HR measures the rate at which the teacher's top-$k$ outputs hit the positive query set.}
\label{tab:teacher-prompt}
\begin{tabularx}{\linewidth}{c X c}
\hline
No. & Prompt Construction & HR \\
\hline
1 & Append the user's clicked \texttt{keyword} as an isolated final \texttt{[SEP]} segment & 0.40775 \\
2 & Insert the clicked \texttt{keyword} into the RAG histories field & 0.39987 \\
3 & Randomly insert all posterior queries into existing fields & \textbf{0.49638} \\
4 & Mask part of the query words before random insertion & 0.49041 \\
\hline
\end{tabularx}
\end{table}

Table~\ref{tab:teacher-prompt} compares different teacher-prompt construction strategies. Appending posterior queries only at the end of the input or injecting them into a fixed field exposes the privileged evidence in a narrow part of the input and yields limited teacher HR. Randomly inserting posterior queries into existing fields performs best, suggesting that distributing the privileged evidence across multiple input locations makes it easier for the teacher to associate posterior queries with the surrounding context. We therefore adopt Strategy 3 (randomized insertion) as the default teacher prompt for PIOPD.

\begin{figure}[!t]
    \centering
    \includegraphics[width=0.85\linewidth]{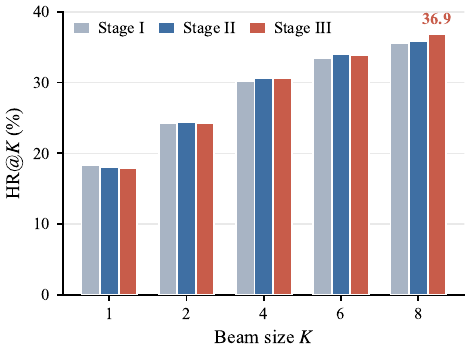}
    \caption{Hit rate across training stages under different beam sizes.}
    \label{fig:hr}
\end{figure}

\noindent\textbf{PIOPD Diagnostics.}
As shown in Figure~\ref{fig:hr}, the gains of PIOPD are mainly reflected at larger cutoff positions rather than HR@1. Compared with Stage I and Stage II, Stage III achieves the best HR@8 while maintaining comparable performance at smaller cutoffs. This indicates that PIOPD improves the coverage of the generated candidate set instead of simply sharpening the top-ranked prediction. By distilling the posterior-aware teacher distribution into the online student, PIOPD transfers soft token-level preferences over multiple semantically plausible query variants. Consequently, beam search can retain more relevant alternatives in the middle and tail ranks, which leads to better recall under multi-candidate generation. This directly explains the offline tradeoff above: OneBar sacrifices a tiny amount of top-1 sharpness for broader candidate coverage.

\begin{table}[tp]
\centering
\caption{Performance of PIOPD objective variants.}
\label{tab:objective-dynamics}
\footnotesize
\setlength{\tabcolsep}{3pt}
\renewcommand{\arraystretch}{1.05}
\begin{tabular}{lcc@{\hspace{0.6cm}}lcc}
\toprule
\multicolumn{3}{c}{\textit{Different KL objectives}} &
\multicolumn{3}{c}{\textit{Top-$k$ support}} \\
\cmidrule(lr){1-3}\cmidrule(lr){4-6}
Variant & HR@8 & MRR@8 &
Variant & HR@8 & MRR@8 \\
\midrule
FKL & 0.3662 & 0.2406 &
Top-16 & 0.3615 & 0.2381 \\
RKL & 0.3626 & 0.2385 &
Top-50 & 0.3612 & 0.2384 \\
JSD & 0.3632 & \textbf{0.2411} &
Top-100 & 0.3611 & 0.2379 \\
CKD & 0.3617 & 0.2385 &
\textbf{Full} & \textbf{0.3690} & \textbf{0.2402} \\
\textbf{FKL+RKL} & \textbf{0.3690} & 0.2402 &
& & \\
\bottomrule
\end{tabular}
\end{table}






We compare several posterior-matching objectives under the same PIOPD protocol. Given a student rollout, the teacher and student are evaluated on the same student-visited prefix at each decoding step, and the loss is computed on on-policy trajectories rather than static label prefixes. The objectives include Forward KL (FKL)~\cite{kullback1951information,hinton2015distilling}, Reverse KL (RKL)~\cite{kullback1951information}, Jensen--Shannon divergence (JSD)~\cite{lin1991divergence}, and Constrained Knowledge Distillation (CKD)~\cite{ni2026star}. For CKD, we adapt its support-constrained matching terms to student rollouts, ensuring a fair comparison under the same on-policy posterior-matching setting.

\begin{figure}[t]
\centering
\includegraphics[width=0.85\linewidth]{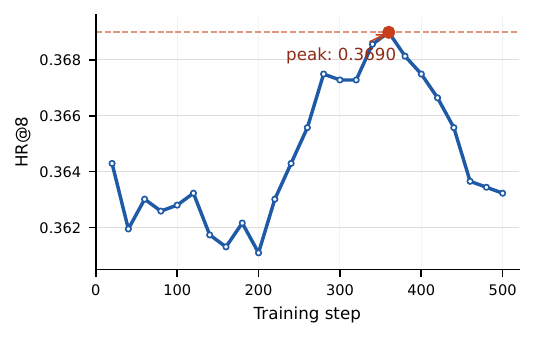}
\caption{Checkpoint-level HR@8 curve of PIOPD training.}
\label{fig:opsd-training-curve}
\end{figure}

\begin{table}[!t]
\centering
\caption{Ablation on Progressive Initialization Stages}
\label{tab:staged-initialization}
\begin{tabularx}{\linewidth}{X c c}
\toprule
Training Strategy & HR@8 & MRR@8 \\
\midrule
Stage I: SFT Only & 0.3564 & 0.2398 \\
SFT $\rightarrow$ PIOPD (Skipping Stage II) & 0.3628 & 0.2397 \\
SFT $\rightarrow$ Stage II (List-wise Alignment Only) & 0.3586 & 0.1789 \\
SFT $\rightarrow$ Stage II $\rightarrow$ PIOPD (Full) & 0.3690 & 0.2402 \\
\bottomrule
\end{tabularx}
\end{table}

Table~\ref{tab:objective-dynamics} reports HR@8 and MRR@8 under different objectives and vocabulary support strategies. Among single-direction objectives, FKL achieves stronger HR@8 than RKL and JSD, suggesting that covering teacher-supported alternatives is crucial for bottom-bar query generation. However, CKD does not yield the best performance. To explain this result, we conduct a token-level distribution analysis by comparing teacher and student output distributions on rollout and gold-label tokens within the top-$k=100$ and top-$m=100$ supports. As shown in Fig.~\ref{fig:entropy-top1-prob-distribution}, the teacher remains more calibrated while the student tends to over-concentrate on top tokens. The teacher--student top-$k$ overlap reaches 69.55\%, while the fraction of gold tokens that fall in the student top-$m$ but outside the teacher top-$k$ is 0\%. This indicates that the student rarely assigns high probability to unsupported tokens. Thus, the main challenge is not suppressing student-exclusive tokens, but preserving valid alternatives with moderate teacher probability. This explains why full-distribution matching with complementary FKL and RKL terms achieves the strongest overall alignment.

For vocabulary support, full-vocabulary PIOPD is more effective in our setting. Recent analyses~\cite{li2026rethinking} suggest that top-$k$ on-policy distillation can approximate full-vocabulary matching in long-horizon reasoning, where a small shared token set often concentrates most probability mass. In contrast, our diagnostics show that the student-side probability mass outside the teacher top-$k$ is only 1.47\%, while the teacher entropy remains substantially higher than the student's (0.9039 vs. 0.6393). This suggests that the teacher assigns meaningful probability to a wider tail of semantically valid query expressions. Truncating the loss to top-16 or top-100 tokens therefore removes critical tail information, whereas full-vocabulary matching preserves posterior knowledge beyond the truncated head, explaining its consistent advantage in Table~\ref{tab:objective-dynamics}.

Fig.~\ref{fig:opsd-training-curve} further shows the PIOPD checkpoint curve.
We find that PIOPD training converges rapidly: although the KL loss decreases monotonically, the student reaches its best downstream performance very early in training, and continued optimization beyond this point leads to a gradual performance drop. This motivates validation-based early stopping rather than loss-based stopping.

Table~\ref{tab:staged-initialization} evaluates progressive initialization. Applying PIOPD directly after Stage I may make distillation depend on weaker student rollouts, introducing noisier token-level posterior targets and slowing convergence. Stage II serves as an intermediate list-wise alignment step, allowing the policy to absorb coarse candidate-level preference orderings before token-level distillation. PIOPD then refines this better-initialized policy with dense token-level soft labels. This supports the progressive design: Stage II provides coarse preference alignment, and Stage III turns it into fine-grained posterior internalization.


Periodically refreshing the teacher with an exponential moving average (EMA) underperforms the frozen-teacher baseline. The frozen teacher achieves 0.3690 HR@8 and 0.2402 MRR@8, compared with 0.3651/0.2394 for EMA-5 and 0.3639/0.2400 for EMA-100. This suggests that, in PIOPD, a stable reference distribution is more beneficial than teacher adaptivity. Updating the teacher during distillation may introduce additional target drift, which can reduce the consistency of the posterior supervision received by the student.

\subsection{Online A/B Testing}

\begin{table}[t]
\centering
\caption{Online quality and business evaluation results. Quality changes are reported as absolute reductions in bad-case rate for OneBar relative to the MCA pipeline, while business metrics are reported as relative improvements over onlineMCA within the eligible video scope.}
\label{tab:online-ab}
\footnotesize
\setlength{\tabcolsep}{3.5pt}
\renewcommand{\arraystretch}{1.05}
\begin{tabularx}{\linewidth}{
>{\raggedright\arraybackslash}X c
>{\raggedright\arraybackslash}X c}
\toprule
\multicolumn{2}{c}{\textit{Manual Quality Evaluation}} &
\multicolumn{2}{c}{\textit{Online A/B Business Metrics}} \\
\cmidrule(lr){1-2}\cmidrule(lr){3-4}
Bad-case Category & Change & Metric & Change \\
\midrule
Overall query-quality bad cases & -9.00 pp & Query Exposure & +16.91\% \\
Video-query irrelevance & -3.66 pp & Query Click & +18.68\% \\
Literal-quality issues & -1.00 pp & Query CTR & +0.19\% \\
Risk-control issues & -4.33 pp & Guided Orders & +20.36\% \\
 &  & Guided GMV & +21.67\% \\
\bottomrule
\end{tabularx}
\end{table}

\definecolor{weakred}{HTML}{FFF1F0}
\definecolor{oursgreen}{HTML}{F0FAF2}
\definecolor{casegray}{HTML}{F5F6F7}
\definecolor{tagblue}{HTML}{EAF2FF}
\definecolor{tagtext}{HTML}{2F5D9B}

\newcolumntype{Y}{>{\raggedright\arraybackslash}X}

\newcommand{\casetag}[1]{%
\begingroup
\setlength{\fboxsep}{1.5pt}%
\colorbox{tagblue}{\textcolor{tagtext}{\scriptsize\textbf{#1}}}%
\endgroup
}

\begin{table*}[t]
\centering
\caption{Representative qualitative cases illustrating how OneBar handles weak, missing, or risky trigger evidence. Queries are translated to English for readability.}
\label{tab:qualitative-case}
\footnotesize
\setlength{\tabcolsep}{5pt}
\renewcommand{\arraystretch}{1.12}

\begin{tabularx}{\textwidth}{@{}Y Y@{}}
\toprule
\rowcolor{casegray}
\textbf{Weak / Incomplete Signal} & \textbf{Recovered Evidence / OneBar Effect} \\
\midrule

\rowcolor{casegray}
\multicolumn{2}{@{}l@{}}{\textbf{Case 1: Risk rejection} \quad \casetag{Reject}} \\
\midrule
\cellcolor{weakred}
\textbf{Trigger.} A family-vlog video shows room tidying and a conversation with a child. The RAG histories only mention the child name with a weak smoking-related cue.

\textbf{Weak signal.} The MCA extension path produces a weakly grounded query related to the child name, with potential harmful-guidance risk as ``Xiaobao Smoking''. 
&
\cellcolor{oursgreen}
\textbf{Recovered decision.} OneBar outputs \texttt{[REJECT]} instead of exposing a bottom-bar query.

\textbf{Effect.} The model avoids over-amplifying a sparse and unsafe association, showing that the generator can learn when not to recommend under insufficient and risky evidence.
\\

\midrule

\rowcolor{casegray}
\multicolumn{2}{@{}l@{}}{\textbf{Case 2: ANN anchor completion} \quad \casetag{Complete}} \\
\midrule
\cellcolor{weakred}
\textbf{Trigger.} The caption mentions ``6800 budget three-cylinder beast'' and contains tags related to CFMOTO 675. The other field is empty, while the clicked target query is ``CFMOTO 675NK''.

\textbf{Missing signal.} The raw caption indicates the broader CFMOTO 675 series but misses the specific \texttt{NK} model suffix.
&
\cellcolor{oursgreen}
\textbf{Recovered evidence.} ANN anchors from behavior-related query pools concentrate on specific variants such as ``CFMOTO 675NK exhaust sound'', ``used price'', ``modified 675NK'', and ``sale price''.

\textbf{Effect.} OneBar completes the missing model suffix and generates the precise query ``CFMOTO 675NK'', rather than stopping at a generic CFMOTO 675 query.
\\

\midrule

\rowcolor{casegray}
\multicolumn{2}{@{}l@{}}{\textbf{Case 3: Complementary retrieval} \quad \casetag{Compose}} \\
\midrule
\cellcolor{weakred}
\textbf{Trigger.} The caption only identifies a spine-surgery doctor and a health-education scenario. The item title, OCR, and RAG histories evidence fields are null.

\textbf{Missing signal.} The target query asks about effective medicine for lumbar spinal stenosis, but the disease entity is absent from the visible trigger evidence. ANN retrieval mostly provides the generic ``what medicine is most effective'' query template with mismatched entities.
&
\cellcolor{oursgreen}
\textbf{Recovered evidence.} User history retrieves nearby disease queries such as ``best medicine for lumbar spinal stenosis''. ANN retrieval further contributes the medication-query template as ``What is the most effective medication for relieving menstrual cramps?''. Finally, OneBar returns ``What is the most effective medication for lumbar spinal stenosis?''

\textbf{Effect.} OneBar combines the correct disease entity with the appropriate medication-seeking template and recovers the target intent.
\\

\bottomrule
\end{tabularx}
\end{table*}

To verify the effectiveness of OneBar in real-world online deployment, we conducted a rigorous online A/A test for 7 days followed by an online A/B test for 8 days in the query recommendation scenario for video-related search in Kuaishou's main single-column feed. The absolute traffic allocation was 9.12\% for MCA and 9.12\% for OneBar. Within the eligible video scope, we compared OneBar against the production onlineMCA baseline. In this scenario, related queries are not exposed for every video, because indiscriminate query exposure would reduce the overall CTR and negatively affect the long-term value of the main Kuaishou feed. Therefore, the primary efficiency objective is to increase the Exposure of video-related queries as much as possible under a CTR constraint, thereby driving more search traffic and ultimately improving orders and GMV. Due to data security concerns, all results are presented in relative values.

Table~\ref{tab:online-ab} summarizes the online business impact. Compared with online MCA, OneBar increases Query Exposure by 16.91\% and Query Click by 18.68\%, while still maintaining a slight CTR gain of 0.19\%. This indicates that the additional exposure does not dilute query attractiveness. The increased search traffic further translates into 20.36\% more guided Orders and 21.67\% higher guided GMV.

In the query recommendation scenario, query quality is also critical to user experience. Specifically, query quality refers to both the relevance between the query and the video, and the literal quality of the query itself; a high-quality query should be free of typos, redundancy, truncation, and risk-control violations. To assess the actual impact on online search experience, we sampled 400 exposed queries from the eligible video scope, including 200 examples from OneBar and 200 from the MCA pipeline, and conducted manual evaluation. The results in Table~\ref{tab:online-ab} show that OneBar improves relevance, literal quality, and risk-control behavior at the same time. To make these error modes concrete, Table~\ref{tab:qualitative-case} presents representative cases covering unsafe weak grounding, ANN-based model completion, and complementary retrieval.

Overall, the online A/B test verifies that OneBar can be deployed synchronously in the main feed and improve both search traffic and downstream conversion under the eligible video scope. The manual evaluation further shows that the gains are not obtained by simply increasing exposure volume; OneBar also reduces bad cases in relevance, surface-form quality, and risk control.

\subsection{Online Deployment Analysis}
In this section, we mainly discuss two questions about the online 
deployment of OneBar for video-related query recommendation and provide
our investigations to facilitate further research.

\textbf{1) What are the main aspects of online gains for OneBar?}
We further group online gains by video publication time. OneBar achieves positive gains across all publication-time buckets, demonstrating the robustness and effectiveness of our method. Notably, the CTR improvement is particularly significant for videos published within the last 30 days. This is because the Collaborative-Multimodal Intent Grounding module enables OneBar to deeply understand video content by integrating textual metadata, multimodal evidence, and collaborative query signals. As a result, OneBar can generate more relevant and timely queries even for recently published videos that have limited behavioral feedback.

\begin{figure}[t]
    \centering
    \includegraphics[width=0.95\linewidth]{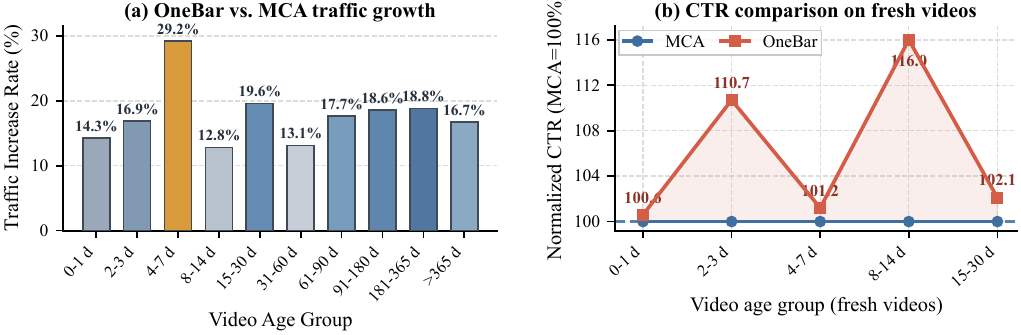}
    \caption{Online gains of OneBar over MCA across video and user groups.}
    \label{fig:mca-onebar-final}
\end{figure}

We further analyze online gains across user activity levels. Based on search frequency, clicked and purchased items, and overall spending, users are divided into five groups, U0--U4, with increasing activity. OneBar achieves statistically reliable gains across all groups, indicating that its benefits are not limited to a specific user segment. For low-activity users, content-grounded evidence, including textual metadata, multimodal understanding, and collaborative query anchors, helps mitigate sparse behavior. For high-activity users, richer behavioral signals and posterior preference learning better capture personalized search intent. These results show that OneBar remains effective across diverse user activity levels.

\begin{figure}[t]
    \centering
    \includegraphics[width=0.95\linewidth]{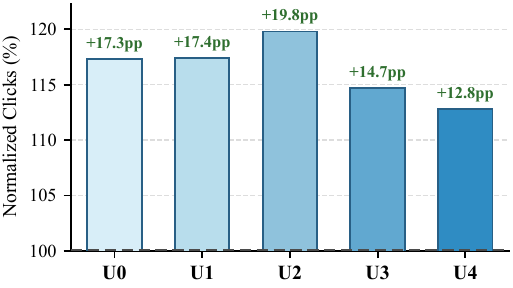}
    \caption{Online gains of OneBar across different user activity levels.}
    \label{fig:buyer-5levels}
\end{figure}

\textbf{2) How does OneBar improve query quality in manual evaluation?}
From the manually annotated bad cases, we observe that OneBar reduces the overall query-quality bad-case rate by 9.00 percentage points, including a 3.66-point reduction in video-query irrelevance and a 1.00-point reduction in literal-quality issues. Further inspection shows that both irrelevance and literal-quality problems are often caused by low-quality video-side information, especially when video titles are noisy or when newly uploaded videos lack behavior-derived query evidence. To mitigate this issue, OneBar incorporates high-quality multimodal evidence and ANN-retrieved query anchors, which effectively improve both query relevance and surface-form quality. In addition, the reduction in risk-control bad cases mainly comes from the rejection-generation strategy, where the model is trained to avoid generating queries for videos that may involve policy-sensitive content such as pornography, gambling, or drug-related topics. As a result, the risk-control bad-case rate is reduced by 4.33\% in manual evaluation.

\section{Conclusion and Future Work}
This paper presented OneBar, an end-to-end generative framework for content-grounded bottom-bar query recommendation. Instead of relying on a multi-stage retrieval cascade, OneBar formulates recommendation as direct generation from the currently consumed video trigger, unifying collaborative-multimodal intent grounding, a compact \texttt{[SEP]}-delimited evidence schema, and progressive preference internalization within a single low-latency generator. In particular, Preference-Internalized On-Policy Distillation (PIOPD) injects posterior behavior signals directly into the policy, which removes the dependence on a separately trained reward model while keeping the serving interface unchanged. Extensive offline evaluations and a large-scale online A/B test on Kuaishou's main feed demonstrate that OneBar consistently improves query quality and business metrics, including substantial gains in exposure, clicks, guided orders, and guided GMV, together with a lower query-quality bad-case rate.

\bibliographystyle{ACM-Reference-Format}
\bibliography{ref}

\end{document}